\documentclass{aa}
\usepackage{graphicx}
\usepackage{txfonts}
\usepackage{hyperref}
\usepackage{amsmath}
\begin{document}

   \title{Investigating the vertical distribution of the disk as a function of radial action: Results from simulations}

   %\subtitle{results from simulations}

   \author{Yunpeng Jia
          \inst{1}\thanks{jiayunpeng11@mails.ucas.ac.cn}
          \and
          Chengqun Yang\inst{2}
          \and
          Yuqin Chen\inst{3,4}
          \and
          Cuihua Du\inst{4}
          \and
          Gang Zhao\inst{3,4}
          }

   \institute{School of Physics, Shangqiu Normal University, Shangqiu, 476000, P. R. China
         \and
         Shanghai Astronomical Observatory, Chinese Academy of Sciences, Shanghai, 200030, P. R. China
         \and
            CAS Key Laboratory of Optical Astronomy, National Astronomical Observatories, 
Chinese Academy of Sciences, Beijing, 100101, P. R. China
\and 
School of Astronomy and Space Science, University of Chinese Academy of Sciences, Beijing 100049, P. R. China
}

   %\date{Received April 15, 2021; accepted May 16, 2021}

% \abstract{}{}{}{}{} 
% 5 {} token are mandatory
 
  \abstract
  {}
   {Previous research has established a relationship between radial action and scale height in Galactic disks, unveiling a correlation between radial and vertical heating. This finding poses a challenge to our existing comprehension of heating theories and consequently encodes crucial insights into the formation and heating history of Galactic disks. In this study, we perform N-body simulations with the aim of verifying the existence of this correlation between radial action and scale height, thereby enhancing our comprehension of the heating history of Galactic disks.
   }
   {We conducted a simulation featuring a disk embedded within a static dark matter halo potential, and systematically analyzed the correlation between radial action and scale height across every snapshot. Furthermore, we augmented this simulation by incorporating massive, long-lasting particles to examine their impact on the aforementioned relationship. 
   }
   {We find that the relationship between radial action and scale height in our simulations can be described by the same functional form observed in previous work. Furthermore, the relationships derived from our simulations align well with those of the Galactic thin disk. However, they do not coincide with the inner thick disk but exhibit a rough correspondence with the outer thick disk, suggesting the possibility that additional heating mechanisms 
   may be required to explain the inner thick disk. We also find that the mean radial action and scale height undergo rapid increases during the initial stages of the simulation, yet remain relatively unchanged as the disk evolves further.
    By tracing example particles, we uncover a correlation between radial and vertical heating in our simulation: as a particle in the disk gains or loses radial action, its vertical motion tends to oscillate on a more or less extended orbit, accompanied by a tendency to migrate outward or inward, respectively. 
   The massive, long-lasting particles in our simulation contribute to disk heating by solely enhancing the rate of increase in scale height with radial action, while maintaining the functional form that describes the relationship between these two variables.
 }
  {We have successfully replicated the functional form previously reported in research, thereby confirming a correlation between radial and vertical heating. This achievement enhances our understanding of heating theories in galactic disks.}


   \titlerunning{Investigation on the scale height as a function of radial action}
   \maketitle
   
%
%-------------------------------------------------------------------

\section{Introduction}
Understanding the mechanisms and consequences of Galactic disk heating is crucial for gaining insights into the formation and evolution of our Galaxy.
Disk heating is a consequence of fluctuations in the gravitational field. The sources of fluctuation include
but are not limited to giant molecular clouds\citep[e.g.,][]{Spitzer1953,Lacey1984,Hanninen2002}, 
transient spiral arms\citep[e.g.,][]{Barbanis1967,Lynden-Bell1972,Carlberg1985}, bars\citep{Saha2010,Grand2016}, minor mergers\citep[e.g.,][]{Quinn1993,House2011},
and fuzzy dark matter halos\citep{Yang2024}.

Heating causes the disk stars' orbits to become more eccentric over time. Quantities commonly utilized to characterize heating are 
velocity dispersions and actions. The radial or vertical action is an integral of the radial or vertical component of a star's momentum over its orbit, 
measuring the extent of radial or vertical oscillations. Stars with low radial action usually have more circular orbits. 
In contrast, stars with higher radial action possess more eccentric orbits.
The advantage of using action to describe heating lies in the fact that action is an adiabatic invariant under any gradual potential changes 
that reflect the Galaxy's growth\citep{Ting2019}. Therefore, it is more appropriate for reflecting heating when studying the secular evolution of a disk, 
although velocity dispersion is easier to measure than actions.
Previous works usually study heating by using age-velocity dispersion relations \citep[e.g.,][]{Nordstrom2004,Holmberg2009,Casagrande2011,Aumer2016,Sanders_Das2018},
or by modeling action-age relations\citep{Ting2019,Garzon2024}.
It is found that a combination of heating from internal 
heating sources\citep[giant molecular clouds, spirals, and bars,][]{Aumer2016}, or giant molecular clouds alone\citep{Ting2019}, can explain the observational results of the overall Galactic thin disk.
Additionally, clues indicate that the inner and outer disks may have experienced different heating histories\citep{Mackereth2019,Garzon2024}.

Theoretical study of heating mechanisms suggests that giant molecular clouds are effective at both radial and vertical heating, while
transient spiral arms are only effective at radial heating as their spatial structures are much larger than the amplitude of oscillations perpendicular to the disk plane\citep[e.g.,][]{binney2008-book}. 
However, the correlation between radial and vertical heating in the Galactic disks remains poorly understood, 
and if such a correlation exists, the precise connection between these two forms of heating remains unclear.

Recently, \citet[][hereafter J23]{jia2023} reported a correlation between radial action, $J_R$, and scale height, $z_0$, for both the Galactic thin and thick disks.
They discovered that the function $z_0=\sqrt{J_R/a}+b$, where $a$ and $b$ are free parameters, describes this correlation well in the case of the thin disk. However, for the thick disk, this function should be used with caution.  Additionally, they found that the relationship between radial action and scale height varies between the inner and outer regions of the thick disk, in contrast to the thin disk where the relationship remains the same between the inner and outer regions.
In the isothermal limit and under the epicycle approximation, the function $z_0=\sqrt{J_R/a}+b$ inherently establishes 
a constant correlation between the radial and vertical velocity dispersion. Specifically, $\sigma_R^2 \propto (\sigma_z/\sqrt{8\pi G\rho_0}-b)^2$, 
because $J_R  \propto \sigma_R^2$ and $\sigma_z^2=8\pi G\rho_0 z_0^2$. Therefore, the relationship between radial action and scale height in J23
reveals a correlation between radial and vertical heating.
These observational results are likely the cumulative effect of various potential heating agents. Heating from giant molecular clouds, 
owing to its effectiveness in both the radial and vertical directions, is inherently suspected to play a pivotal role in shaping this relationship, 
whereas transient spirals are likely to contribute minimally. Nevertheless, our analysis in a simplified N-body simulation reveals that, 
regardless of whether massive, long-lasting particles, which to some extent represent giant molecular clouds, are included or not, a disk embedded within a static dark matter halo potential is capable of reproducing the functional form of the relationship between radial action and scale height, as is reported in J23.

This paper is structured as follows. In Section 2, we briefly describe the simulation. The results are presented in Section 3. Finally, a discussion and conclusions are given in Section 4.

%--------------------------------------------------------------------
\section{Simulations}
As a first attempt to directly study the correlation between radial and vertical heating in simulations through
an investigation of the relationship between radial action and scale height, we
carried out a simple simulation with a disk embedded in a dark matter halo. The simulation was run
with the help of the fourth version of GADGET; that is, GADGET-4 \citep{Springel2021}.
For practical reasons, a live dark matter halo is often modeled with
particles more massive than disk particles, which may introduce spurious heating that depends on the mass resolution of dark matter \citep{Ludlow2021}.
Therefore, to enhance computational efficiency and ensure a tightly controlled experimental environment, we opted to utilize 
 a fixed dark halo potential instead of a live dark matter halo.

The initial condition (IC) of the model was generated using a Python code: galstep\footnote{https://github.com/ruggiero/galstep}, which consists of a dark matter halo and a stellar disk.
The dark matter halo follows a Hernquist density profile\citep{Hernquist1990}:
\begin{equation}
	\rho(r) =\frac{M}{2\pi} \frac{a}{r} \frac{1}{(r+a)^3}.
\end{equation}
We chose the mass $M=10^{12}$ $\mathrm{M}_\odot$ and the scale factor $a=47$ kpc, which are same as those in \citet{Ruggiero2017}. 
The disk follows an exponential density profile: 
\begin{equation}
\rho(R,z)=\frac{M_b}{8\pi z_0{h_R}^2} \mathrm{sech}^2(\frac{z}{2z_0}) \mathrm{exp}(-\frac{R}{h_R}).
\end{equation}
We set the mass $M_b=5\times 10^{10}$ $\mathrm{M}_\odot$, a scale-length $h_R=2.5$ kpc, and the scale height $z_0=0.15$ kpc.
The IC contains $5\times10^6$ disk particles, each of which has a mass of $10^{4}$ $\mathrm{M}_\odot$.

\begin{figure*}
	\includegraphics[width=\hsize]{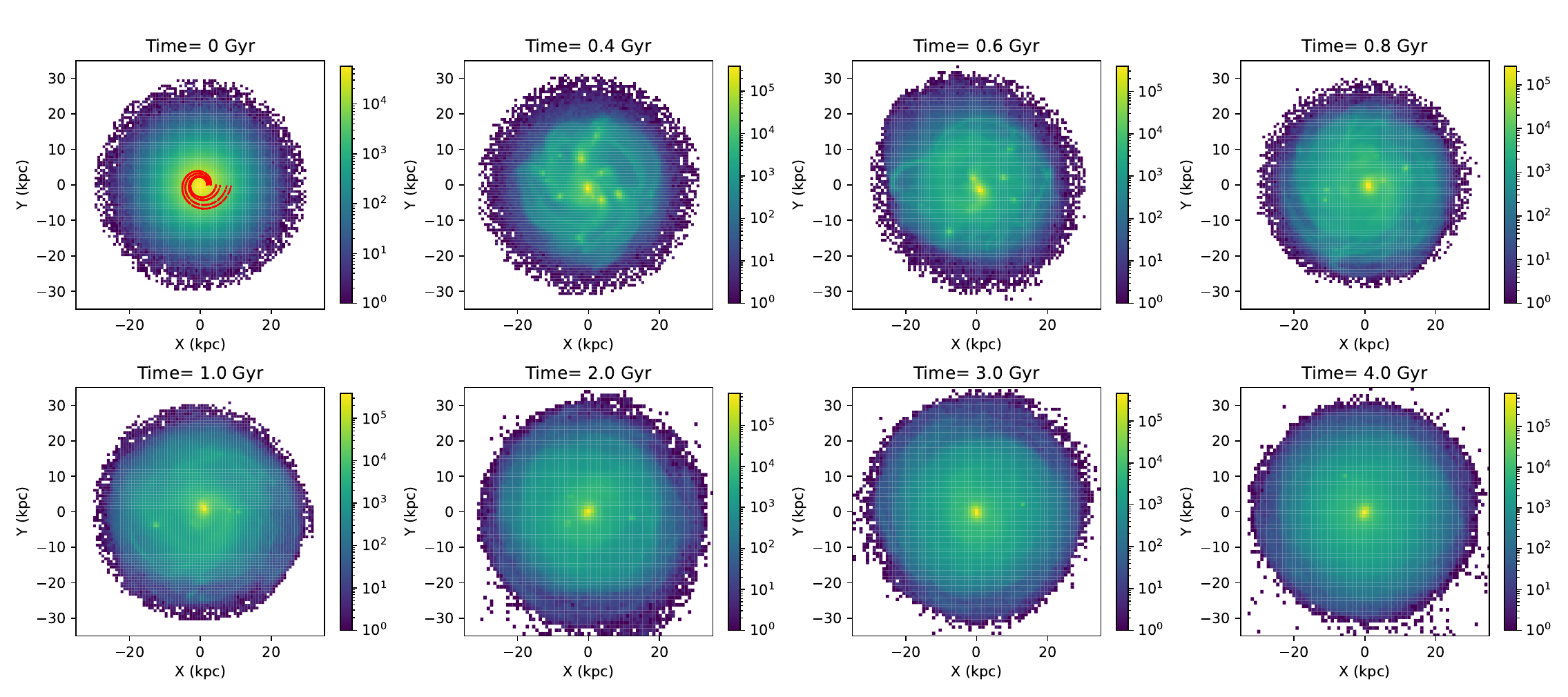}
	\caption{Snapshots of the disk from the face-on view. The upper left panel specifically depicts the distribution of massive, long-lasting particles within the IC (indicated by red dots), arranged in distinct spiral-like patterns. The snapshots presented in the other panels are derived from 
		the ``without-GMC'' simulation (a simulation that does not include massive, long-lasting particles).
 \label{GMC}}
\end{figure*}

To understand disk heating caused by massive objects, such as giant molecular clouds, we conducted an additional simulation that introduces long-lived, large-mass particles into the disk within the IC mentioned above. However, given that giant molecular clouds are typically short-lived, this simplified simulation may not fully replicate their impact on a real galaxy. Nonetheless, it provides us with valuable insights into this complex problem.
As giant molecular clouds are primarily located in the spiral arms, the massive particles are been placed on spiral arm in the top left panel of Fig.~\ref{GMC}, following the work of \citet{Kokaia2019}.
These logarithmic spiral arms were fit by \citet{Hou-Han2014} from observations of massive star-forming regions and giant molecular clouds.
The massive particles were introduced on perfect circular orbits, with their rotational velocities determined by $\sqrt{a_R(R)R}$, where $a_R(R)$
is the radial gravitational acceleration. The masses of all massive particles were assigned with $10^6$ $\mathrm{M}_\odot$, which approaches those of giant molecular clouds. 
The IC contains 1000 massive particles, which contribute $10^9$ $\mathrm{M}_\odot$ to the final disk. 

We applied a fixed gravitational softening length of 30 pc for all particles, which is the same as that applied to baryonic particles in  \citet{Aumer2016}.
We refer to the first simulation, which comprises a disk and a static dark halo, as the ``without-GMC'' simulation, and the second simulation, 
which incorporates massive particles in addition to the components of the first simulation, as the ``GMC'' simulation.
Some snapshots of the without-GMC simulation are illustrated in Fig.~\ref{GMC}.
The irregularities in the disk emerge during the early epoch of the simulation and gradually fade over time, until finally the disk becomes axis-symmetric.
These irregularities, some of which look like spiral structures, could be responsible for the disk heating, and we show that the disk is heated during precisely this period of the simulation.

\section{Results} 
\begin{figure*}
	\centering
	\includegraphics[width=\hsize]{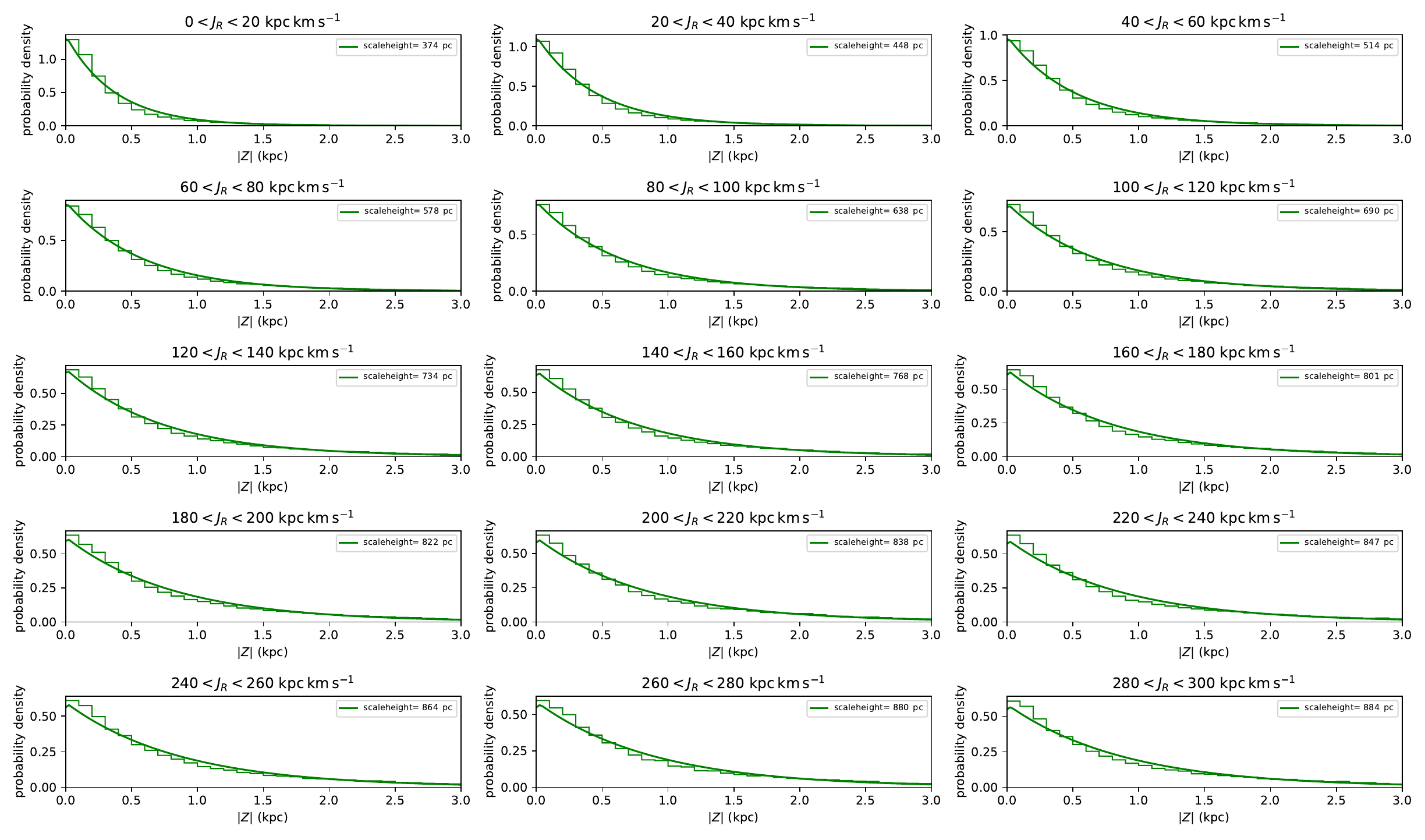}
	\caption{Vertical distributions of disk particles in GMC simulations at 4 Gyr. 
		The solid lines in each panel denote the best-fit density profiles, $\rho(z) \propto \mathrm{exp}(-z/z_0)$. The best-fit scale heights, $z_0$, are listed in the legends.  \label{z-distributions}}
\end{figure*}

\begin{figure*}
	\centering
	\includegraphics[width=\hsize]{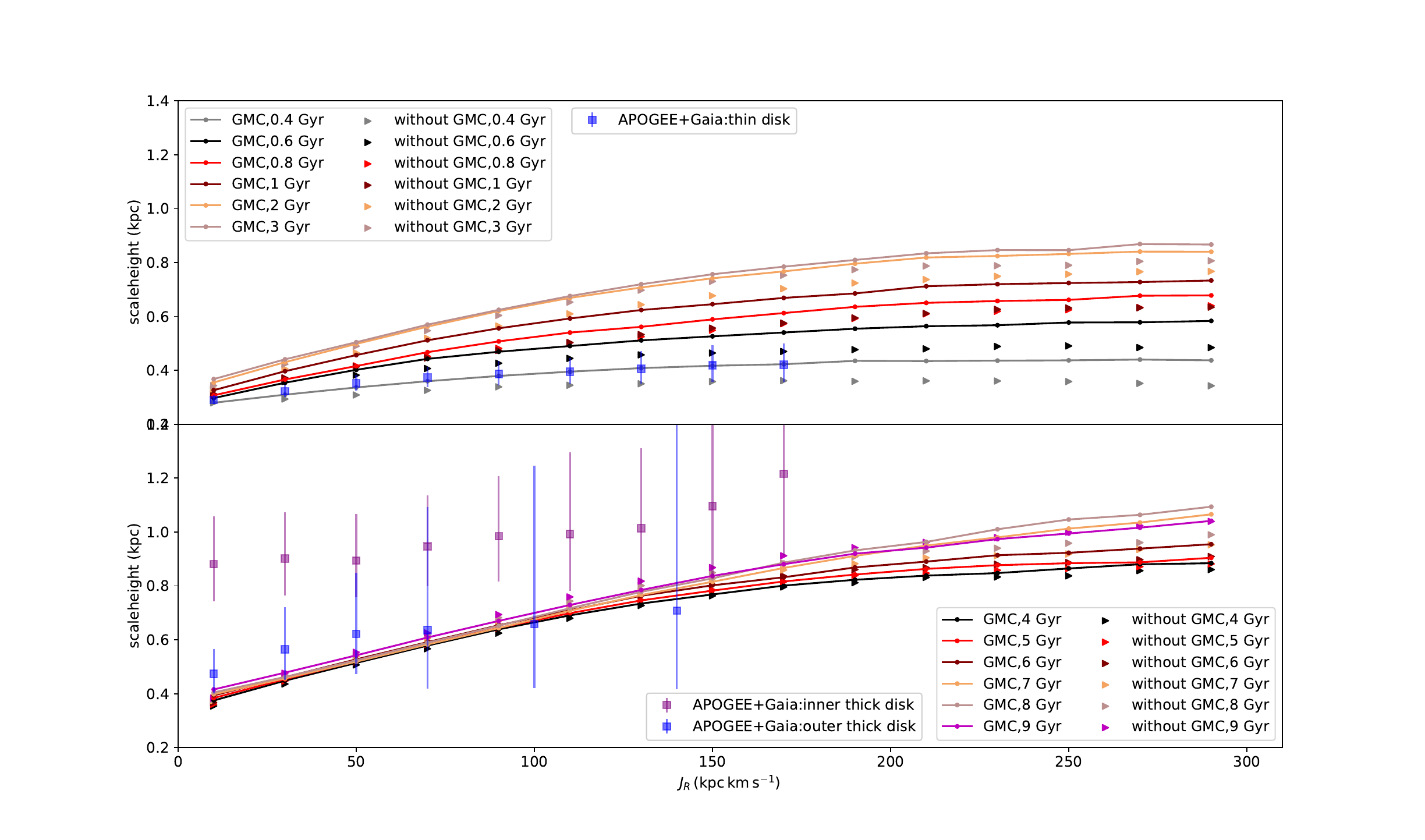}
	\caption{Relationships between radial action and scale height for GMC and without-GMC simulations. The lines with dots show the results of the GMC simulation, while the triangles show the results of the without-GMC simulation. To more clearly demonstrate that the relationships between radial action and scale height 
		in the GMC and without-GMC simulations are nearly identical after 4 Gyr, the upper panel exclusively displays results from simulations up to 3 Gyr, while the results from simulations beyond 3 Gyr are presented in the bottom panel. In each panel, the lines and triangles are colored differently to represent various simulation time points. If a line and its corresponding triangles share the same color, they indicate the same simulation time. For clarity, the typical range of uncertainties in scale height, which is between 0.01 kpc and 0.02 kpc, is not shown here.
		The squares accompanied by error bars represent the results of J23, which were obtained using data from APOGEE and Gaia.  \label{jr-h}}
\end{figure*}

For the purpose of investigating the relationship between radial action and scale height in the disk, we calculated the radial actions of disk particles using AGAMA \citep{Vasiliev2019AGAMA} and estimated the scale heights of the disk particles (excluding massive, long-lasting particles) in radial action bins for each snapshot of the simulations.
In order to facilitate a comparison with the observations in J23, the density profile used to estimate the scale height, $z_0$, is 
$\rho(z) \propto \mathrm{exp}(-z/z_0)$, which is different to that in the IC. The estimated scale height for this density profile in the IC is 207 pc.
The scale heights were estimated with the maximum likelihood technique \citep{Bienayme1987}.
The uncertainties of scale heights were estimated using the same method as in J23. Specifically, 
we calculated the likelihood 500 times by employing the observed data and the simulations of the best-fitting model, with individual Poisson noise applied. 
The resulting likelihood range determines the uncertainty. Typically, the uncertainty of the scale height ranges from 0.01 kpc to 0.02 kpc. 
However, in the high radial action bins ($J_R>260$ $\rm kpc\, km\, s^{-1}$), it is around 0.03 kpc.
As an example, we arbitrarily chose 4 Gyr to illustrate the vertical distributions within radial action bins for the GMC simulations in Fig.~\ref{z-distributions}.
It is evident that the vertical distributions of the disk particles can be properly described by this density profile across all radial action bins, and this consistency holds true for every snapshot taken from all of our simulations.
The resulting relationships between radial action and scale height are illustrated in Fig.~\ref{jr-h}.
The relationships obtained from GMC and without-GMC simulations are shown by solid lines and triangles, respectively.
The colors used for the lines and triangles are the same for the same simulation time. 
For comparison, the relationships that observed by J23 are also shown in this figure, which is marked by the squares. 
In J23, the inner and outer thin disk gives a similar relationship with
the overall thin disk; therefore, the relationship of the overall thin disk is only presented in this figure.

Before summarizing the key findings from Fig.~\ref{jr-h}, it should be pointed out that for the without-GMC simulation, 
the scale heights do not exhibit a monotonic increase with radial actions prior to 0.4 Gyr (about one rotation time). 
Additionally, the AGAMA package unfortunately failed to calculate
radial action for the GMC simulation before 0.4 Gyr. Therefore, we have chosen to present the relationships starting from 0.4 Gyr in this figure.

From Fig.~\ref{jr-h}, we find that:

1) After 0.4 Gyr, the relationships between radial action and scale height in all the snapshots can be consistently described using the same functional form that was proposed in J23; that is, $z_0=\sqrt{J_R/a}+b$, where $a$ and $b$ are free parameters. While we have chosen not to include the best-fit lines for this functional forms for clarity, the perfect agreement between the data points of the thin disk observed by J23 (depicted by blue squares in the top panel) and the GMC simulation results at 0.4 Gyr provides compelling evidence for the validity of the proposed functional relationships. 

2) The scale heights gradually reach saturation over time, with this saturation process occurring more rapidly for smaller radial actions. 
The saturation level roughly matches the one of the outer thick disk observed by J23, but is notably lower than the one of the inner thick disk also observed by J23.

3) During the first 2$\sim$3 Gyr, the rates of increase in scale height with radial action in GMC simulations  exceed those in without-GMC simulations. 
However, these rates gradually converge over time, becoming virtually identical by 4 Gyr. Subsequently, the differences in the relationships between simulations with and without GMCs become negligible.

As is observed in J23, we also find that the distributions of radial actions for all the snapshots of our simulations can be approximately described by pseudo-isothermal distributions, $J_R\propto \exp(-J_R/\hat{J_R})$, where $\hat{J_R}$ is a free parameter that depends on the snapshot and stands for the temperature of the disk. 
In principle, this free parameter should match the mean radial action of the snapshot, given the characteristic of pseudo-isothermal distributions. We have verified that in our simulations, the best-fit value of $\hat{J_R}$ indeed aligns with the mean radial action of the snapshots. Therefore, we shall use the mean radial actions to represent the temperature of the snapshots from now on.

\begin{figure}
	\centering
	\includegraphics[width=\hsize]{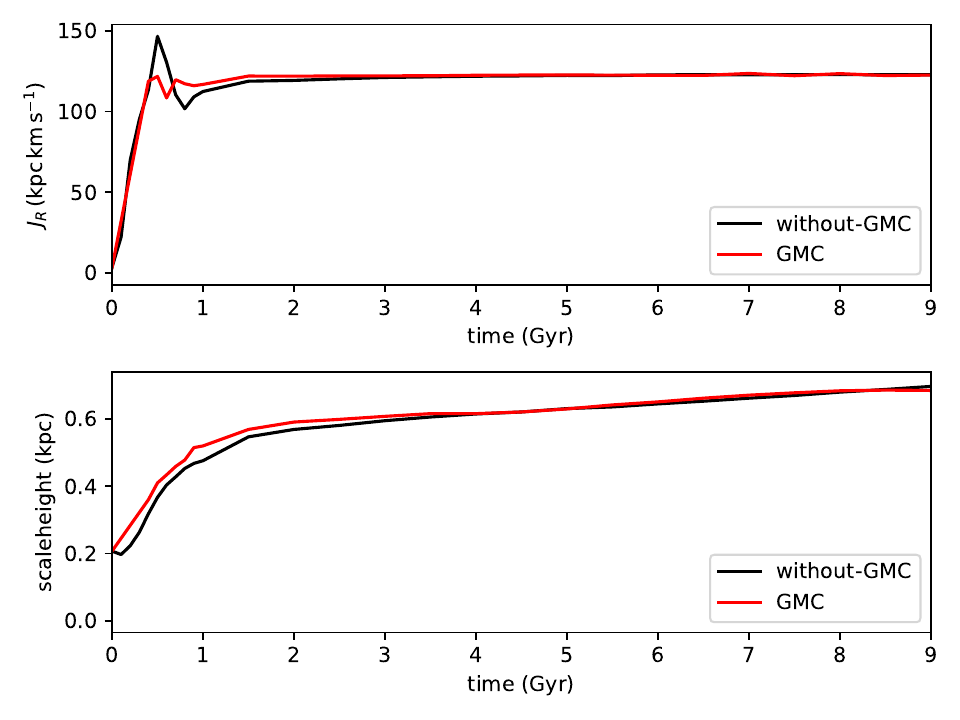}
	\caption{Evolution of the mean radial action (top panel) and scale height (bottom panel) in our simulations.  \label{jr-h-time}}
\end{figure}

In the top panel of Fig.~\ref{jr-h-time}, we present the temporal evolution of the mean radial action, starting from the inception of the simulation runs, 
for both the GMC and without-GMC simulations. 
To directly compare the radial and vertical heating, we also estimated a scale height for each snapshot and then plotted the temporal evolution of scale height, which is shown 
in the bottom panel of Fig.~\ref{jr-h-time}.
Notably, both the mean radial action and the scale height exhibit rapid growth during the initial 1 Gyr of simulation time in both simulations. 
However, a divergence emerges subsequently: the mean radial action attains a plateau beyond this point, whereas the rate of increase in the scale height slows down considerably around the same time point and eventually tends toward zero over time.
While the mean radial actions remain generally identical across both simulations, there is a small difference in the scale heights,
with the GMC simulations exhibiting a generally larger scale height compared to the without-GMC simulations prior to 2$\sim$3 Gyr. However, this difference in scale height evens out after this period, indicating that the inclusion of massive particles in the simulations only enhances heating in the vertical extent. This vertical effect is transient, progressively weakening as time progresses.

\section{Discussion and conclusions}
\begin{figure}
	\centering
	\includegraphics[width=\hsize]{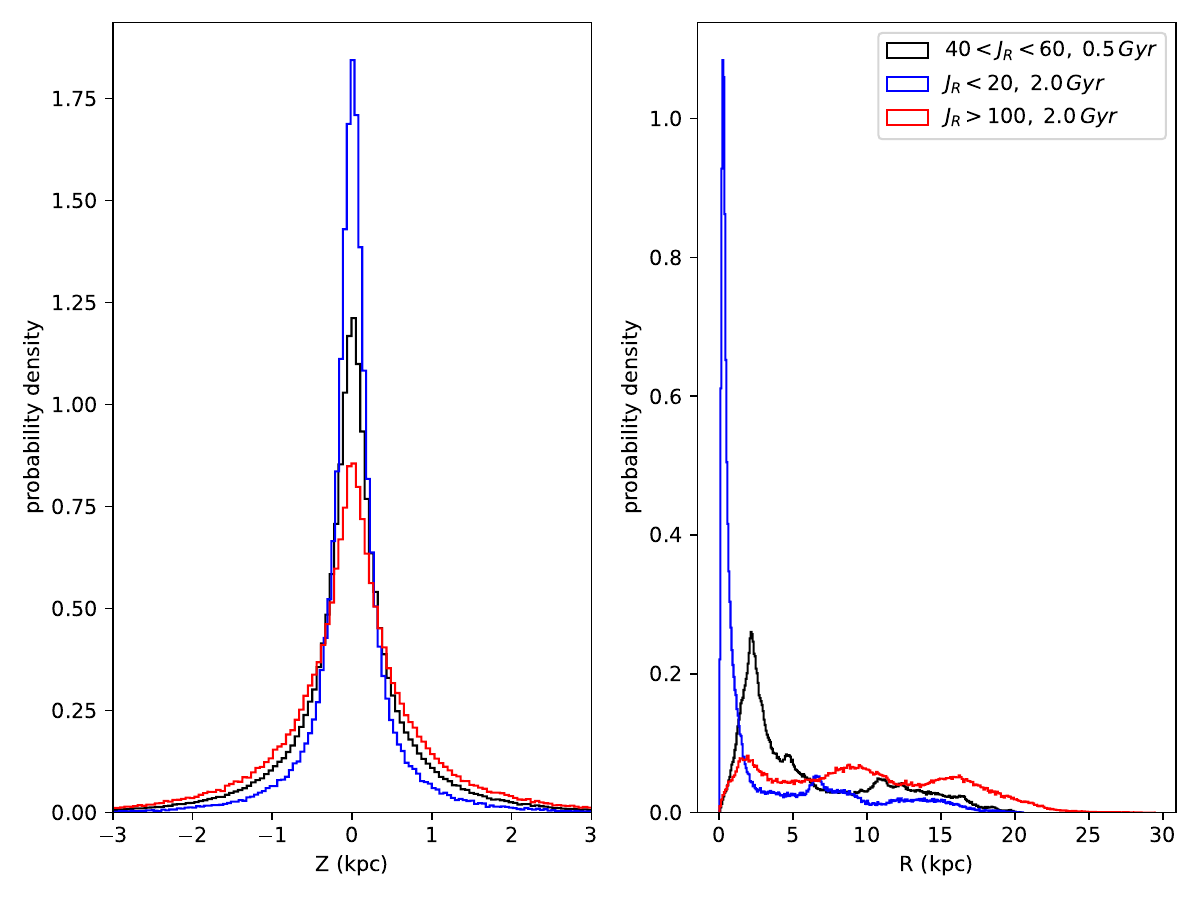}
	\caption{Tracing a population of disk particles with radial action within the range of 40 to 60 $\rm kpc\, km\, s^{-1}$ in the without-GMC simulation from 0.5 Gyr to 2 Gyr. 
	A group of particles with $J_R<20$ $\rm kpc\, km\, s^{-1}$ indicates a loss of radial action, whereas a group of particles with $J_R>100$ $\rm kpc\, km\, s^{-1}$ 
	 indicates a gain of radial action. The vertical and radial distributions are shown in the left and right panels, respectively.
	 \label{tracing}}
\end{figure}

By simulating a disk in a fixed spherical dark halo potential (the without-GMC simulation), we have successfully 
replicated the functional form that has been reported in J23, 
which characterizes the relationship between the radial action ($J_R$) and the scale height ($z_0$) of the disk. 
This relationship is expressed as $z_0=\sqrt{J_R/a}+b$, where $a$ and $b$ are free parameters that depend on the snapshots. 
We also find that the mean radial action and scale height increases rapidly with time during the early epoch of the simulation.
In the meantime, irregular structures (some of which look like spiral structures, see Fig.~\ref{GMC}) are prominent 
in the early epoch of this simulation ($\sim$1 Gyr) .
The above results reveal that the disk is heated by these irregularities, both in the radial and vertical directions, during the early stage of the simulation.
Moreover, the heated disk maintains the aforementioned functional form throughout its entire evolution (more accurately, after $\sim$0.4 Gyr).
Therefore, the radial and vertical heating correlate in such a way that when a particle in the disk gains or loses radial action,
its vertical motion tends to oscillate on a more or less extended orbit, respectively.
To exemplify this intricate interplay, we arbitrarily tracked disk particles exhibiting radial actions within the range of
40 to 60 $\rm kpc\, km\, s^{-1}$. We initiated our analysis at a without-GMC simulation time of 0.5 Gyr, and continued it until 2 Gyr. 
The reason for tracing particles to this time point is that heating is more prominent during the first 2 Gyr (see Fig.~\ref{jr-h-time}). 
However, we have verified that when tracing particles until the end of the simulation (9 Gyr), the following results also hold.
Fig.~\ref{tracing} vividly illustrates that particles that have undergone radial heating -- in other words, those that have gained radial action -- exhibit a larger scale height and a more extended radial distribution compared to particles that started with identical conditions but retained their original radial action. Conversely, particles that were radially cooled display the opposite behavior.
Fig.~\ref{tracing} not only provides clear verification of the aforementioned statements but also reveals that a disk particle undergoing an increase or decrease in radial action tends to migrate radially outward or inward, respectively.

When adding massive, long-lasting particles to the IC of above simulation (the GMC simulation), the functional form of the relationship between radial action and scale height also holds. 
However, there is a higher rate of increase in scale height with radial action during the first 2$\sim$3 Gyr of the GMC simulation, compared to the without-GMC simulation.
After that period, the relationships are nearly identical to those obtained in the without-GMC simulation.
In short, heating from these massive, long-lasting particles is only prominent in the first 2$\sim$3 Gyr in 
increasing the rate of scale height with radial action.
By comparing the temporal evolution of the mean radial action and scale height between the GMC and without-GMC simulations in Fig.~\ref{jr-h-time},
we have uncovered that, at least under the ICs employed in our simulations, incorporating massive particles in the without-GMC simulation  solely enhances vertical heating and accelerates the attainment of saturation in heating.

Upon quantitatively comparing the simulated results with those documented in J23 (see Fig.~\ref{jr-h}), it is observed that the
relationships derived from the simulations closely match the Galactic thin disk. 
They notably fall below the inner thick disk but may instead show a rough correspondence with the outer thick disk. 
This suggests that the heating processes examined in this work are insufficient to explain the inner thick disk. 
As a result, it is plausible that additional heating mechanisms, such as those induced by mergers \citep[e.g.,][]{Quinn1993,Villalobos2008}, may be necessary to account for the inner thick disk. However, we cannot discount the possibility that the inner thick disk was born hot \citep[e.g.,][]{Brook2004,Brook2005,Bournaud2009}.

It is crucial to recognize the limitations inherent in this work, notably that the massive, long-lived particles employed 
do not fully encapsulate the complexities of giant molecular clouds. Moreover, we acknowledge the absence of an 
in-depth analysis of how our findings are shaped by the selection of ICs. 
 While certain implications can be anticipated, such as the fact that the heating rates of massive particles are primarily governed by their mass, local density, and velocity dispersion \citep{Lacey1984, Ludlow2021}, altering these properties may lead to variations in both radial and vertical heating rates.
 Consequently, it is vital to meticulously assess the extent to which the relationship between radial action and scale height is contingent upon the specific ICs chosen, which underscores a pivotal direction for our future research endeavors.

\begin{acknowledgements}
We are grateful to the anonymous referee for the careful reading and constructive suggestions that have clarified this paper. 
This work is supported by the Scientific Research Startup Foundation of Shangqiu Normal University(No. 700155), the Support for Key Research Projects Plan of
Higher Education Institutions in Henan Province(Project No. 25A160001), the International Partnership Program of Chinese Academy of Sciences 
under grant No. 178GJHZ2022040GC,the National Natural Science Foundation of China under grant Nos. 11988101, 12373020 and 12403025, 
and the National Key R\&D Program of China under grant No. 2023YFE0107800.

\end{acknowledgements}

\bibliographystyle{aa} % style aa.bst
\bibliography{ref} % your references Yourfile.bib

\end{document}